%% file: ms.tex
\documentclass[sigconf, 9pt]{acmart}

\newif\ifnotnewacm
\notnewacmfalse

\newif\ifheadnice
\headnicefalse

\input{preamble}

\newif\ifeg
\egfalse

\renewcommand\footnotetextcopyrightpermission[1]{} % removes copyright information

\begin{document}
\title{Technical Report: Implementation of Single Packet Number Space in Multi-Path QUIC}
\author{Yingqi Tang, Yunfei Ma, Yanmei Liu}

\affiliation{Alibaba Group}

\renewcommand{\shortauthors}{Y. Tang, Y. Ma, Y. Liu}

\input{abstract}

\maketitle

\input{intro}
\input{spns}

\input{eval}

\input{discuss}
\input{concl}

%\balance
\begin{acks}
We want to thank Christian Huitema, Mirja Küehlewind, Quentin De Coninck and Olivier Bonaventure for helpful discussions on the implementation details and tradeoffs of SPNS. 
\end{acks}
% \ifeg
% TBD
% \else
% \fi

\newpage
\bibliographystyle{unsrt} % abbrv
\bibliography{ms}
\end{document}

%% file: preamble.tex
\begin{CJK}{UTF8}{gbsn}

\newtheorem{algorithm}{算法}
\renewcommand{\abstractname}{摘要}  % 将Abstract改为摘要
\renewcommand{\refname}{参考文献}   % 将References改为参考文献

\title{***}
\author{***footnote{电子邮件:**}\\[2ex]
*** \\[2ex]
}
\date{20XX年X月}

\maketitle

\tableofcontents
\newpage
在此输入正文，中英文均可。

\songti 
\fangsong 
\biaosong
\heiti
\kaishu

\end{CJK}

%% file: abstract.tex
\begin{abstract}

Over the past few of years, we have witnessed increasing interests in the use cases of multipath QUIC from both industry and academia~\cite{xlink, de2017multipath, draft-ietf-quic-multipath-02}. For example, Alibaba deployed XLINK~\cite{xlink}, a QoE-driven multipath QUIC solution, in Taobao short video and showed benefits in both reduced tail latency and video rebuffering. For the time being, the multipath QUIC protocol is in the process of standardization at the IETF QUIC working group, with the draft recently updated to version 02~\cite{draft-ietf-quic-multipath-02}. The focus of the draft is to provide basic guidance on the implementation so that we can encourage more exploration, testing, and finally, an accelerated adoption of this technology. However, draft-02 has brought up an open issue on whether the multipath QUIC should be implemented using single packet number space (SPNS) or multiple packet number space (MPNS), as in the current draft, both options co-exist. 

Knowing that one cannot draw a solid conclusion without experiments, we implement both SPNS and MPNS at Alibaba and measured their performance. The goal is to help the community better understand the implication, and we hope this report can be a useful resource for engineers and researchers who are interested in deploying multipath QUIC. We also suggest readers to read \cite{de2022packet}, which is another comparative study between MPNS and SPNS based on Picoquic~\cite{picoquic}.

\end{abstract}

%% file: intro.tex
%!TEX root = sigconf.tex

\section{Introduction}\label{sec:intro}

We assume the reader is familar with QUIC. For the backgroud knowledge, please refer to RFC9000~\cite{rfc9000}. There is also a fantastic blog~\cite{lucasquic} by Lucas Pardue on this topic that we encourage readers to read. Our today's discussion will be on SPNS vs. MPNS. So what are they?

\noindent
\textbf{SPNS: }
Single Packet Number Space (SPNS) means that in multipath QUIC, packets transmitted on different paths will be assigned packet numbers from the same packet number space. For example, if packet number $\#1$ is assigned to a packet on the first path, the same number (i.e., $\#1$) cannot be assigned to a packet on another path. In other words, packet number overlapping is not allowed.

\noindent
\textbf{MPNS: }
Multiple Packet Number Space (MPNS) means that in multipath QUIC, packet numbers are assigned based on a per-path manner, and packets on different paths use separate packet number spaces. For example, two paths can both number their packets with the number $\#1$, $\#2$, and $\#3$. Packet number overlapping is perfectly fine.

There are pros and cons in each option. For SPNS, the benefits include: (1) it does not require any change in how the nonce is computed for AEAD; (2) it allows multi-path transport even when zero-CID is used. The drawbacks of SPNS include: (1) it cannot straightforwardly use the algorithms in RFC9002~\cite{rfc9002} for loss detection, RTT sampling, and recovery; (2) the packet ACK size is larger due to increased number of holes in the packet range records, and (3) there is a lack of ECN support.

For MPNS, the benefits include: (1) the loss detection and recovery are more straightforward and easy to implement; (2) the ACK size is not bloated as the packet number is continuous on each path; (3) supporting ECN is easy with path identifiers.  The drawbacks of MPNS are: (1) it requires the change of how we compute the nonce; (2) it needs the CID to identify each path, and thus, MPNS does not work with zero-CID cases.

Indeed, the qualitative properties such as the nonce modification, the support of zero-CID cases, and the support of per-path ECN are easy to reach concensus among the working group. However, questions like the implementation complexity and performance tradeoff typically do not have a yes-or-no answer without further experiments and real-world deployments. 

Based on the draft-ietf-quic-multipath-01~\cite{draft-ietf-quic-multipath-01}, we implement multipath QUIC for both SPNS and MPNS. An endpoint can decide whether to use single path, multipath with MPNS or multipath with SPNS through the negotiation with a peer. For SPNS, we highlight the algorithm changes required for both sender and receiver to maintain performance, including how to obtain accurate and timely RTT estimations. Our implementation is based on XQUIC~\cite{xquic}, an open source QUIC library in C language developed by Alibaba Group.

Our experiement consists of two parts: (1) controlled experiment with Mahi-mahi mpshell emulation tools and (2) a large-scale A/B test in Taobao RPC over HTTP scenario with over 1.5 million request samples. In this typical scenario, our results showed that SPNS had a slight performance degradation of less than $1\%$ on average request completion time compared with that of MPNS. We observed that its impact on the request completion time for small-sized RPC requests was not significant. However, SPNS became slower than MPNS when the request size was large. In terms of the acknowledgement cost, if not constrained, the average size of ACK frame of SPNS could go $50\%$ higher than that of MPNS.

We tried to reduce the ACK frame size of SPNS using strategies discussed on the WG draft's Github~\cite{quicwg-mpquic-repo}. Our results showed that we were able to bring down average ACK frame size of SPNS by $40\%$. However, in doing so, we observed decrease in transmission efficiency.

%% file: spns.tex
\section{Single Packet Number Space}\label{sec:spns}

\begin{figure}[tb]
	\centering
	%\vspace{-1mm}
	\includegraphics[width=0.9\linewidth]{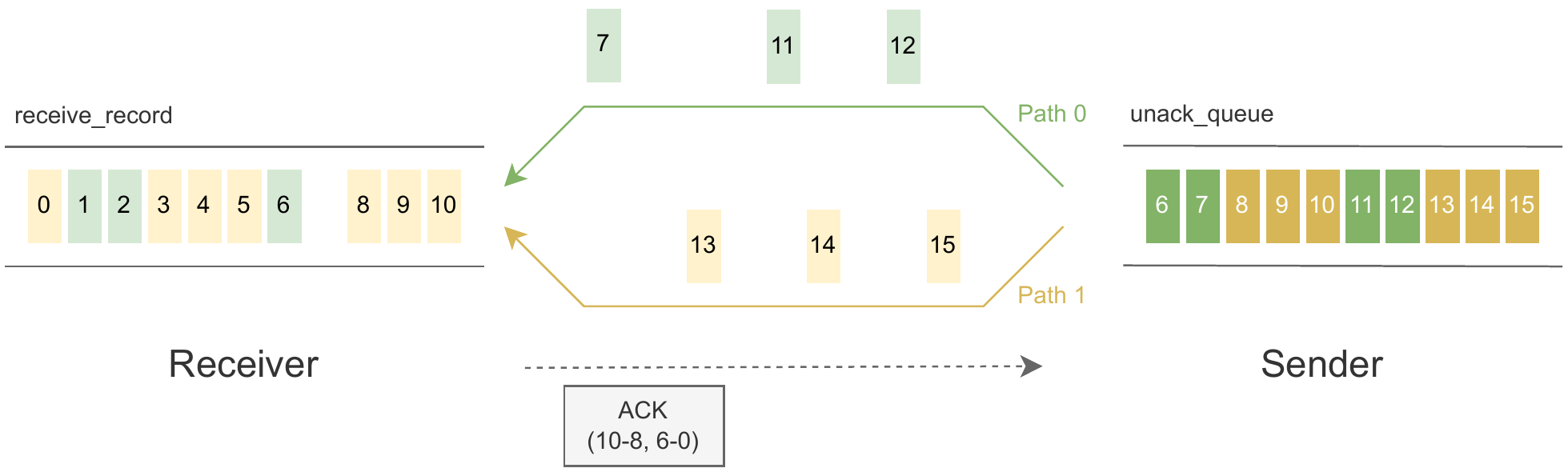}
	\vspace{0mm}
	\caption{\small{Sending and receiving data in SPNS.}} 
        \label{fig:spns:context-origin}
	\vspace{-5mm}
\end{figure}

Conceptually, a packet number space is the context in which a packet can be processed and acknowledged. For single path QUIC, the packet numbers in one space forms a continuous sequence starting from 0. Such a continuity simplifies algorithms like loss recovery and RTT measurements. However, this assumption is no longer valid in multipath QUIC using SPNS. Fig.~\ref{fig:spns:context-origin} illustrates a possible scenario. There are two paths in the connection: Path 0 and Path 1. The sender has sent packets with packet numbers $\{1,2,6,7,11,12\}$ at Path 0 and $\{0,3,4,5,8,9,10,13,14,15\}$ at Path 1. The receiver has received packets $\{1,2,6\}$ at Path 0 and $\{0,3,4,5,8,9,10\}$ at Path 1. What ends up happening is that the packet numbers are spread across multiple paths, which yields a discontinuous sequence on each path. Such a discontinuity leads to multiple problems:

\begin{itemize}[leftmargin=*]
    \item  Packets reordering. Due to heterogeneity of different paths, the packets may arrive out of order.  The reordering leaves holes in the received packet number record and causes ACK range to bloat. If the number of ACK ranges is larger than the maximum count specified by the endpoint, some of them cannot be sent, which causes a sender to incorrectly consider packets loss, even though they have been actually delivered.
    \item  Change in the loss detection algorithm. Due to the packet number discontinuity, the sender needs to maintain an association between sent packet numbers and the path over which these packets were sent, for the purpose of loss detection and congestion control. This introduces additional complexity in the implementation.
    \item  Change in the RTT estimation algorithm. An RTT sample in SPNS may be the sum of the uplink delay of one path and downlink delay of another path. In the current WG draft, it is undefined how this RTT sample should be used to compute the smoothed round-trip time estimation. On the other hand, it is difficult to control the RTT update frequency of each path, which requires the receiver to design specific strategies.
\end{itemize}

Next, we discuss how we implement SPNS's loss detection and RTT measurment algorithms.

\begin{figure}[tb]
	\centering
	%\vspace{-1mm}
	\includegraphics[width=0.9\linewidth]{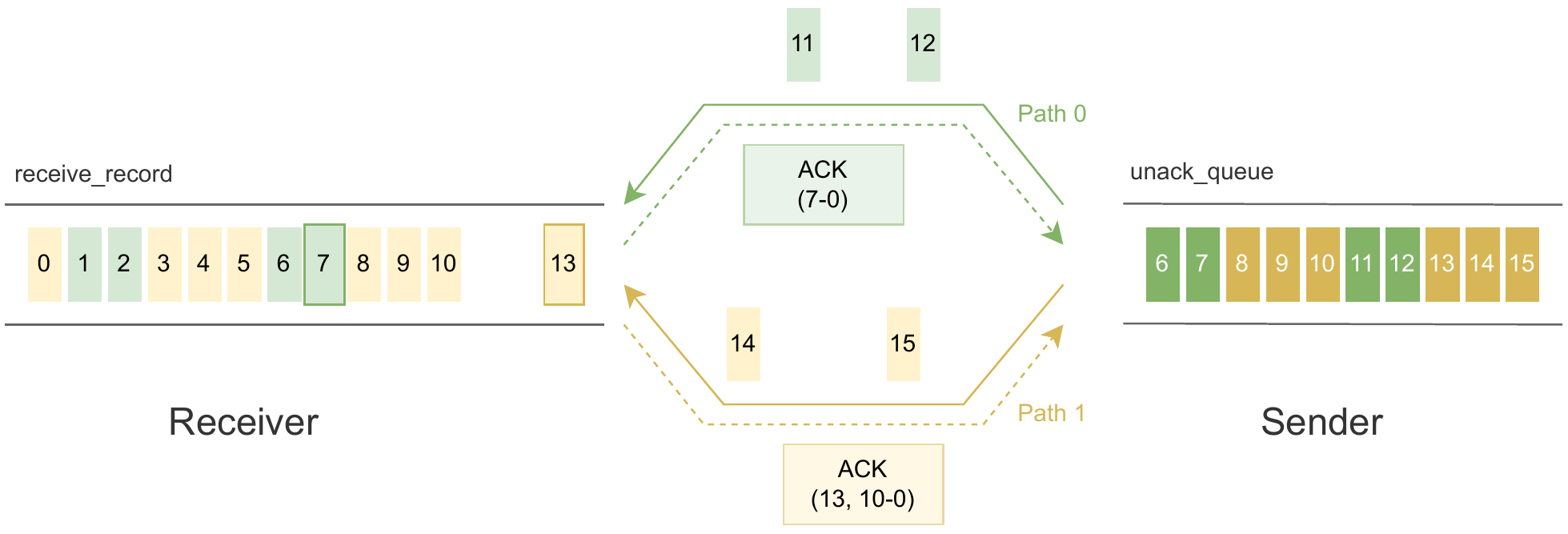}
	\vspace{0mm}
	\caption{\small{Write ACK frame based on the largest received packet of the path.}} 
        \label{fig:spns:context-opt}
	\vspace{-5mm}
\end{figure}

\subsection{Loss Detection}\label{sec:spns:loss}

Regarding the loss detection, we followed the basic strategies that has been defined in the current working group draft for SPNS, including:

\begin{itemize}[leftmargin=*]
    \item  Maintain Packet Context. The sender stores an index association between each path and the packets it sends.
    \item  Loss Detection. In single path QUIC, an unacknowledged packet is declared lost if it was sent kPacketThreshold packets before an acknowledged packet, or it was sent long enough in the past. Instead of simply using (largest acknowledged PN - kPacketThreshold) , we track the sending history of a path to determine which packet was the k-th one before the largest acknowledged packet in the same path.
\end{itemize}

\begin{figure*}[t]
\hspace{0cm} 
  \centering
  \begin{minipage}{.3\linewidth}
  \subfloat[\small{Downlink Throughput of 2 paths}]{
  \includegraphics[width=1\columnwidth, angle=0]{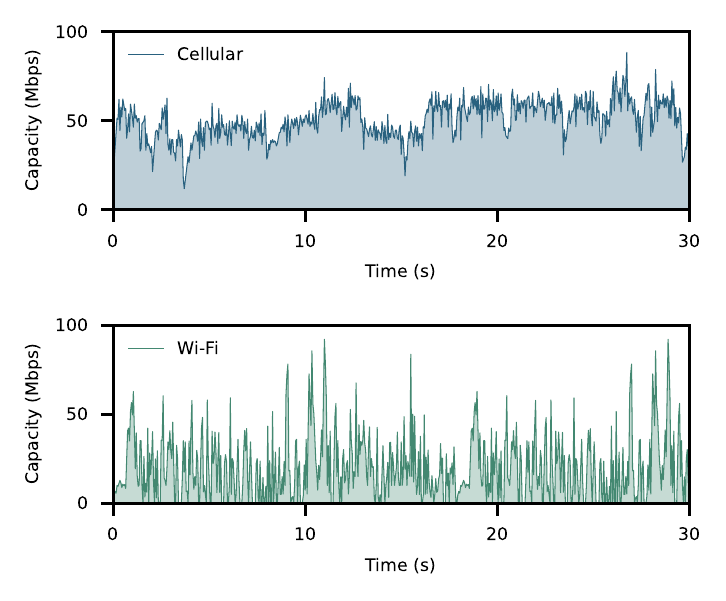}
  \label{fig:spns:capacity}
  }
  \end{minipage}
  \hfill
  \begin{minipage}{.3\linewidth}
  \subfloat[\small{RTT Trace and SRTT of Paths}]{
  \includegraphics[width=1\columnwidth, angle=0]{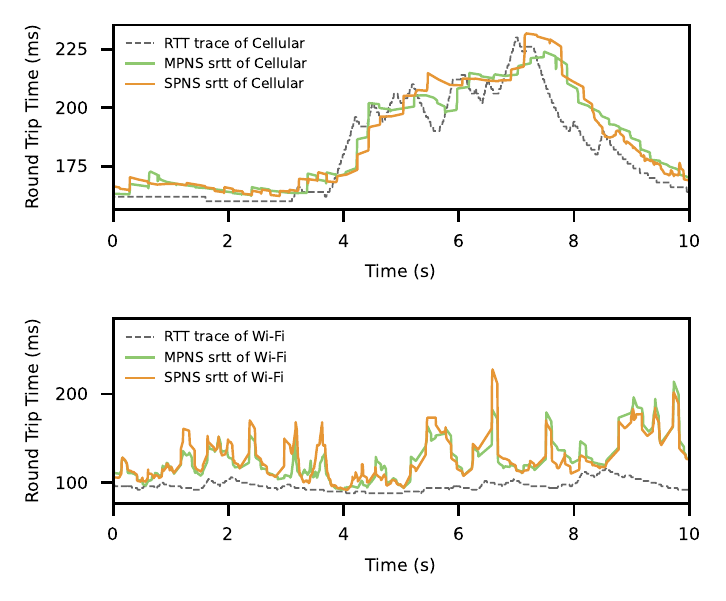}
  \label{fig:spns:rtt}
  }
  \end{minipage}
  \hfill
  \begin{minipage}{.3\linewidth}
  \subfloat[\small{Received Packet Number}]{
  \includegraphics[width=1\columnwidth]{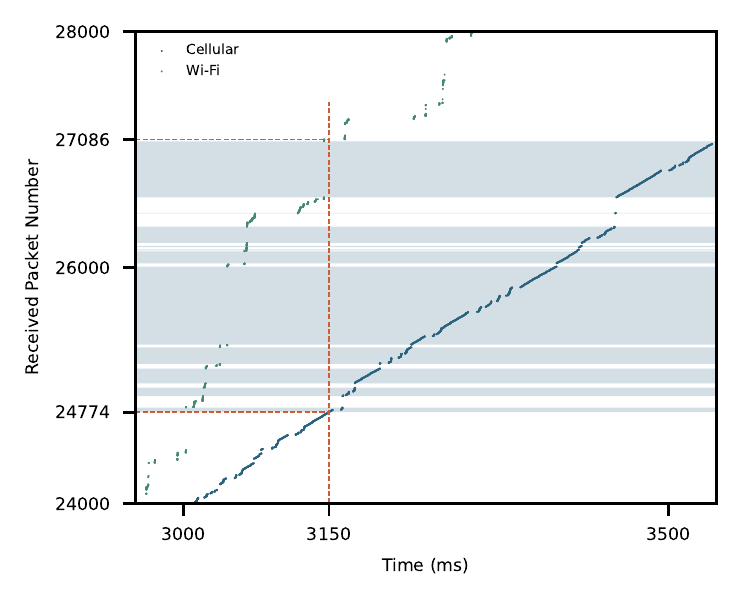}
  \label{fig:spns:recpn}
  }
  \end{minipage}
  \hfill
 \vspace{0.1in}
  \caption{\small{Controlled experiment. (a) The capacity trace of Cellular and Wi-Fi replayed by Mahi-mahi. (b) The RTT trace of Cellular and Wi-Fi replayed by Mahi-mahi and the smooth RTT of the two paths estimated by the server. (c) The received packet number by the client with SPNS.}}
   \vspace{-3mm}
  \label{fig:spns:emul}
\end{figure*}
% \vspace{-5mm}

\begin{figure*}[t]
\hspace{0cm} 
  \centering
  \hfill
  \begin{minipage}{.35\linewidth}
  \subfloat[\small{CDF of ACK Range Count}]{
  \includegraphics[width=.85\columnwidth, angle=0]{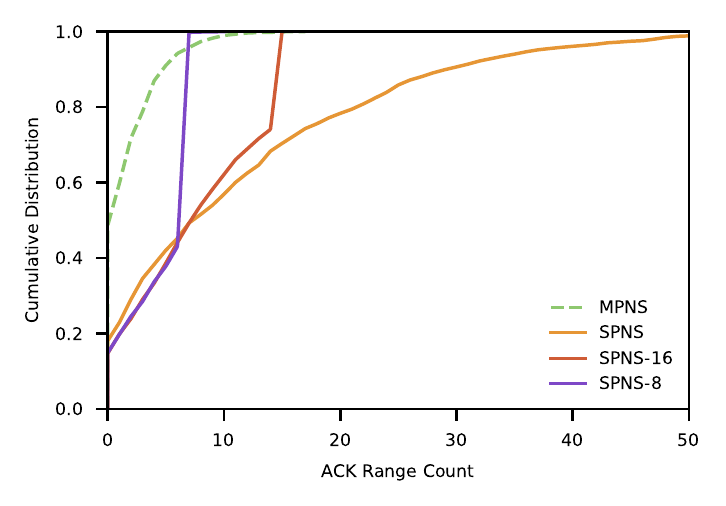}
  \label{fig:spns:ackrange}
  }
  \end{minipage}
  \hfill
  \begin{minipage}{.35\linewidth}
  \subfloat[\small{Transmission Speed Comparison}]{
  \includegraphics[width=.85\columnwidth, angle=0]{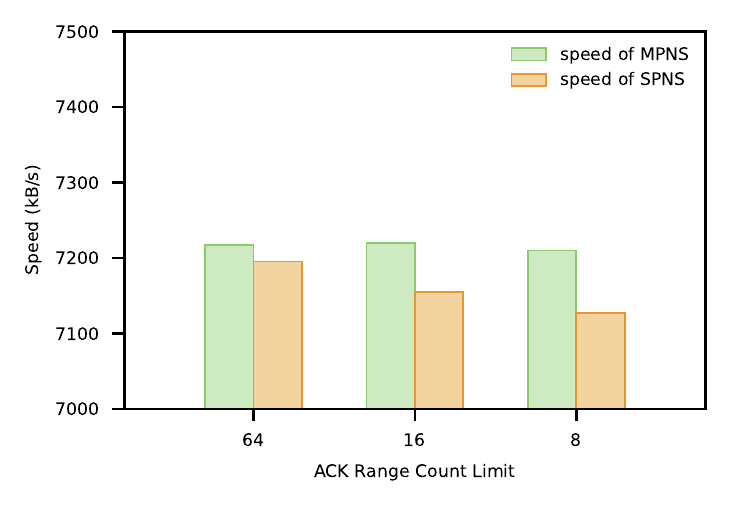}
  \label{fig:spns:emul-speed}
  }
  \end{minipage}
  \hfill
  \vspace{0.1in}
  \caption{\small{Controlled experiment with ACK optimized. (a) The cumulative distribution of ACK range count. (b) The comparison of transmission speed between SPNS and MPNS when tuning the default ACK range count.}}
   \vspace{-3mm}
  \label{fig:spns:ack}
\end{figure*}
% \vspace{-5mm}

\subsection{RTT Estimation}\label{sec:spns:rtt}

According to QUIC RFC9002~\cite{rfc9002}, an endpoint generates an RTT sample on receiving an ACK frame that meets the following two conditions:

\begin{itemize}[leftmargin=*]
    \item  the largest acknowledged packet number is newly acknowledged, and
    \item  at least one of the newly acknowledged packets was ack-eliciting.
\end{itemize}

The RTT sample, $latest\_rtt$, is generated as the time elapsed since the largest acknowledged packet was sent:

\centerline{$latest\_rtt$ = $ack\_time$ - $send\_time\_of\_largest\_acked$}

However, when applying the above algorithm, we encountered the following problems:

\begin{itemize}[leftmargin=*]

    \item  \textbf{RTT of some paths are not updated timely.}
    According to draft-ietf-quic-ack-frequency~\cite{draft-ietf-quic-ack-frequency-02}, the receiver should send an ACK after receiving ack-eliciting-threshold number packets and an ack-eliciting packet must be acknowledged within MAX\_ACK\\\_DELAY. For example, one maintains an ack-eliciting threshold of two and an ACK timer in the connection layer, and pushes the packet into the sending queue after writing ACK frame. The sending path of an ACK packet is not decided before it is actually sent. As a consequence, transmission frequency of ACK on each path is later decided by the multi-path scheduling algorithm. In particular, the number of ACK packets sent by each path was evenly distributed when using Round Robin scheduling, but when using minRTT scheduler, all of the ACKs are sent from the path of shorter delay, which is not useful to the RTT update for other paths.
    
    \underline{Our solution:} Therefore, we specify that the receiver should send back an ACK on the same path which has just received two ack-eliciting packets. In this way, the ACK frequency of a path depends on its received packet count rather than the specific path scheduling algorithm.

    \item  \textbf{ACK frames depend on the largest received packet of the connection, not the path.}
    According to RFC9002~\cite{rfc9002}, the ack range in an ACK frame should start from the largest packet number received in the packet number space and the "ACK delay" field is filled in the time difference between the ACK writing time and the receiving time of the largest received packet. The sender should generate an RTT sample for a path when an ACK is received at that path and the RTT sample conditions are met.
 
    There are two problems with this scheme in practice:

    \begin{itemize}[leftmargin=*]
    
        \item  The created RTT sample may be the sum of the uplink delay of a path and the downlink delay of another path, resulting in inaccurate path RTT estimation. There may be inherent delay difference between paths, leading to the fact that almost all the largest received packets in the ACKs were sent by the path with shortest delay.
    
        \item  On the other hand, even if the receiver sends ACKs evenly on each path, an ACK packet $\#0$ sent first from Path 0 may arrive later than the ACK packet $\#1$ sent after from Path 1, because the delay of Path 0 is larger than that of Path 1. Once the ACK packet $\#1$ of Path 1 is received and processed to generate an RTT sample, the ACK packet $\#0$ which arrives later would be of no use any more.
    
    \end{itemize}

    \underline{Our solution:} Therefore, we specify a new rule: the ACK frame sent on a path should start from the largest received packet number of that path. The sender should maintain an unacked list in each path to avoid no RTT update for those paths with longer delay.

\end{itemize}

Putting it together, \textbf{we implement RTT estimation with the following rules.}

\begin{itemize}[leftmargin=*]
    \item  For packet receiver (ACK sender) :

    \begin{itemize}[leftmargin=*]

        \item  Maintain an ack threshold and an ack timer for each path. A path should send an ACK when it receives two ack-eliciting packets. And an ack-eliciting packet must be acknowledged within MAX\_ACK\_DELAY.

        \item  Write ACK frame based on the largest received packet of the path. The ACK ranges should start with the largest received packet number of that path, which means that the "Largest Acknowledged" field should be the path's largest packet number and the "ACK Delay" field should be the delay time of the path's largest received packet.

    \end{itemize}

    \item  For packet sender (ACK receiver) :

    \begin{itemize}[leftmargin=*]

        \item  Maintain an unacked list for each path to retrieve the packets that has been sent when an ACK is received. It can coexist with the unacked list of the connection layer or packet number space layer.

        \item  Generate RTT sample for a path when the following conditions are met:

        \begin{itemize}[leftmargin=*]

            \item  the largest acknowledged packet number is newly acknowledged by the ACK received from \textbf{this path}, and

            \item  at least one of the newly acknowledged packets was ack-eliciting.
        
        \end{itemize}
        
    \end{itemize}

\end{itemize}

\underline{Example:} Take Fig.~\ref{fig:spns:context-opt} as an example for further illustration, which represents the scenario some time later than Fig.~\ref{fig:spns:context-origin}. We assume that packet $\#7$ arrives before packet $\#13$. When the packet $\#7$ arrives at Path 0, an ACK contained ack ranges $\{(7,0)\}$ should be generated and sent from Path 0. When the packet $\#13$ arrives at Path 1, an ACK contained ack ranges $\{(13), (10,0)\}$ should be generated and sent from Path 1. 

Furthermore, the above rules are also fully applicable for MPNS. The packet number space of each path are independent with each other, so the largest received packet of a path is exactly the largest one of the corresponding packet number space.

\subsection{Controlled Experiments: emulation in Heterogeneous 2-Path Network}\label{sec:spns:emul}

In this section, we first conduct controlled experiment to verify our implementation of SPNS loss detection, recovery and RTT estimation algorithms mentioned in the previous sections. We use Mahi-mahi~\cite{mahimahi} emulation tool that can replay a pair of Wi-Fi \& Cellular traces collected in our office. Throughput of the two path is shown in Fig.~\ref{fig:spns:capacity}. A GET request with a 50MB response is initiated in the emulated 2-path environment, and we choose Cubic as the congestion control algorithm and minRTT as the path scheduling algorithm. Table.~\ref{tab:trans-speed} reports the comparison of the transmission rate and average ACK frame size between MPNS and SPNS, which shows that: (1) the transmission rate of SPNS is slightly lower than that of MPNS; (2) the average size of ACK frame is much larger.

\begin{table}[h]
\vspace{-2mm}
\caption{Transmission Speed (MPNS vs. SPNS)}
\label{tab:trans-speed}
\vspace{-2mm}
\begin{tabular}{|c|c|c|}
\hline
\small  & \small Speed (kB/s) & \small ACK frame size (Byte) \\
\hline
\small MPNS & \small 7217.6 & \small 16.9 \\
\hline
\small SPNS & \small 7195.6  & \small 38.35 \\
\hline
\small Rate & \small -0.30\% & \small +126.92\% \\
\hline
\end{tabular}
\vspace{-4mm}
\end{table}

\subsubsection{RTT measurement accuracy}

Fig.~\ref{fig:spns:rtt} shows a comparison between the replayed delay of emulation tool and the smooth RTT of the two path in the server. From the fourth second, the path delay of Cellular increases significantly and the server in SPNS can perceive the change in a timely and accurate manner, which is the same as that in MPNS. The smooth RTT of Wi-Fi is larger than the replayed delay because Wi-Fi has a lower latency than Cellular such that the endpoint with minRTT scheduler are more likely to send packets from Wi-Fi until the congestion window is filled. The congestion on the Wi-Fi causes a large queuing delay, which makes the actual delay larger than the set value. We verify that our RTT estimation strategy is effective, for that it produces similar results to MPNS.

\subsubsection{Observed Reordering}

Fig.~\ref{fig:spns:recpn} shows an excerpt of the relationship between the packet number received by the client and time in SPNS.  It is not hard to tell the difference of path latency leads to a serious reordering. At the point of 3150ms, the largest received packet number of Wi-Fi is 27086 and the one of Cellular is 24774. The shaded blocks represent the holes in received packet number ranges and there are 11 blocks at this moment. In other word, even if no packet has loss, there are 11 holes in the ACK range from the client due to path heterogeneity, and the number of these holes in the connection is unpredictable.

The reordering problem is more clearly reflected in the cumulative distribution function (CDF) of the ack range count in an ACK frame shown in Fig.~\ref{fig:spns:ackrange}. The ack range count in SPNS is higher than that in MPNS.

\subsection{ACK Frame Size Suppression}\label{sec:spns:ack}

In this section, we try to explore some methods to reduce the ack ranges in SPNS. We believe that there are two reasons for the increasing of ACK range:

\begin{itemize}[leftmargin=*]

    \item  Unpredictable reordering. The emulation results in the previous section show that the path heterogeneity will lead to unexpected reordering and make the number of ack ranges increase rapidly.

    \item  Aggregation of loss. Because multiple paths in a connection share the same packet number space, the packet loss that may occur in each path is aggregated and reflected in the ACK Frame.

\end{itemize}

Based on the recommendation in the draft, we attempt to set the following rules for the receiver (ACK sender) to control the ACK frame size of SPNS:

\begin{itemize}[leftmargin=*]

    \item  Tune the ACK threshold for the out-of-order packets.
    We deviate from the practice that an endpoint SHOULD generate and send an ACK frame without delay when it receives an ack-eliciting packet that is out of order. Instead, the endpoint determines whether to send an ACK according to the ack threshold and the ack timer maintained by each path when receiving an ack-eliciting packet, regardless of whether it is out of order.

    \item  Set \textbf{two} limits for the total number of ranges in an ACK frame. 
    The objective is that the receiver could limit the number of ACK ranges in an ACK frame, but not missing newly received packets that cause spurious losses. To achieve this goal, we specify two limits on count of ACK ranges called the Default\_Limit and the Maximum\_Limit. The ACK range count in an ACK frame SHOULD not exceed the Default\_Limit unless it cannot cover a newly received packet and it MUST not exceed Maximum\_Limit in any case. In this paper, what we tune in the ACK range suppression experiment is the Default\_Limit and the Maximum\_Limit is always set to 64.
\end{itemize}

\subsubsection{Verification}

In this section, we discuss our observations on the ACK size optimization with the same controlled experimental settings in section~\ref{sec:spns:emul}. Fig.~\ref{fig:spns:emul-speed} shows the comparison of the transmission rates. The speed of SPNS degrades as the limit of ACK range count is reduced. As shown in Fig.~\ref{fig:spns:ackrange}, the ack range count of SPNS after optimization decreases significantly, but it is still far from MPNS.

%% file: eval.tex
\begin{figure*}[t]
\hspace{0cm} 
  \centering
  \hfill
  \begin{minipage}{.46\linewidth}
  \subfloat[\small{Speed Histogram for Different Data Size}]{
  \includegraphics[width=.85\columnwidth, angle=0]{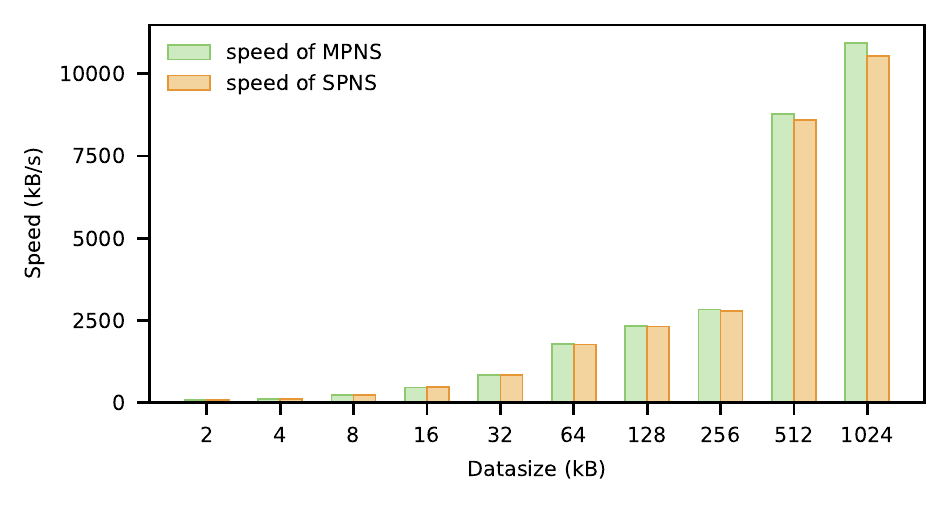}
  \label{fig:abtest:speed}
  }
  \end{minipage}
  \hfill
  \begin{minipage}{.35\linewidth}
  \subfloat[\small{CDF of ACK frame size}]{
  \includegraphics[width=.85\columnwidth, angle=0]{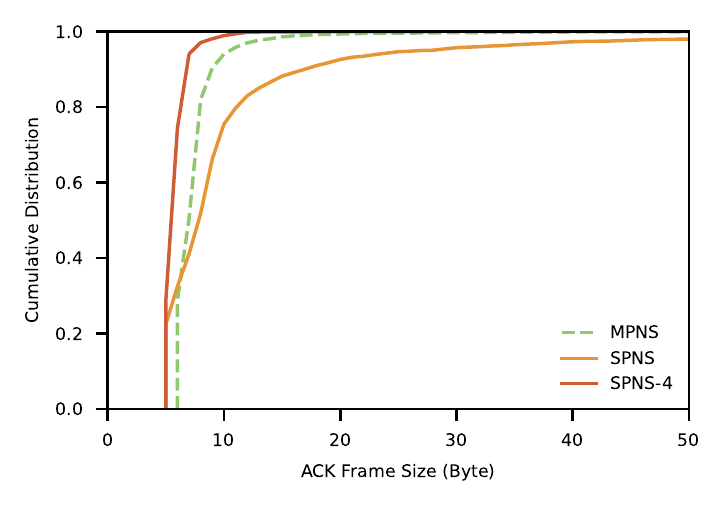}
  \label{fig:abtest:ackframe}
  }
  \end{minipage}
  \hfill
 \vspace{0.1in}
  \caption{\small{A/B Test in Real-world Cases: (a) The average transmission speed of different data size ranges. (b) The cumulative distribution of ACK frame size. }}
   \vspace{-3mm}
  \label{fig:abtest}
\end{figure*}
% \vspace{-5mm}

\section{A/B Test in Real-world Cases}\label{sec:abtest}

In this section, we report the results from a real-world A/B test based on Taobao shopping RPC over HTTP, which mainly includes short connections and parallel small data requests. This type of service have two core metrics:

\begin{itemize}[leftmargin=*]

    \item  Network time. The average value of time spent on the transmission when a request was completed. It is used for quantifying the transmission efficiency.

    \item  One-second completion rate. The ratio of the request completed in one second among all requests that occured. It is used for quantifying the user experience because one second is usually considered as the boundary value at a user's service interruption.

\end{itemize}

In the A/B test, Cubic is used for congestion control and minRTT is used for path scheduling. We first experimented with the basic implementation mentioned in section~\ref{sec:spns:loss} and \ref{sec:spns:rtt}. Afterwards, we experimented with ACK optimization version discussed in section~\ref{sec:spns:ack}. Our data set consists of 1.5 million request samples. Table~\ref{tab:network-time} and Table~\ref{tab:ack-opt} present a comparison between MPNS and SPNS with the field of network time, one-second completion rate, and average ACK Frame size.

\subsection{Multiple PNS vs. Single PNS}

\begin{table}[h]
\centering
\vspace{-2mm}
\caption{Network Time and 1-second Completion Rate}
\label{tab:network-time}
\vspace{-2mm}
\begin{tabular}{|c|c|c|c|}
\hline
\small  & \small Network Time   & \small 1-second           & \small ACK frame size\\
\small  & \small (ms)           & \small Completion Rate    & \small (Byte)\\
\hline
\small MPNS &\small 98.74 & \small 99.1\% & \small 8.97\\
\hline
\small SPNS &\small 99.53 & \small 99.1\% & \small 11.99\\
\hline
\small Rate &\small +0.80\% & \small 0 & \small +33.67\%\\
\hline
\end{tabular}
\vspace{-2mm}
\end{table}

The overall results are summarized in Table~\ref{tab:network-time}. Regarding the basic implementation, the average network time of SPNS increased slightly of $0.8\%$ compared to MPNS. However, the difference on the one-second completion rate was negligible. 

On the other hand, we observed some differences when looking at the data in more details. Fig.~\ref{fig:abtest:speed} shows the speed histogram of difference data sizes, where we calculated the average speed in ranges, which are divided as [2,4), [4,8), [8,16), ..., $[1024,+\infty)$. The ticker '2' indicates the range [2,4) and the bars on ticker '2' refers to the average speed of the requests among 2 to 4 KB. When the data size was larger than 256 KB, the transmission speed of SPNS started to deteriorate significantly compared to MPNS. In other words, the slowdown of SPNS were gradually reflected with the increase of the data size of transmission.

In terms of acknowledgement cost, the value of ACK frame size was small due to the small amount of data. The average size of ACK frame in SPNS was 12 Bytes, which translated to about 4.5 ACK ranges. The average ACK frame size in MPNS was 9 Bytes, which was about 2.5 ACK ranges. The ACK frame of SPNS was $33\%$ larger than that of MPNS.

\subsection{Effects of ACK Frame Suppression}

\begin{table}[h]
\centering
\vspace{-2mm}
\caption{Network Time and 1-second Completion Rate after ACK Suppression}
\label{tab:ack-opt}
\vspace{-2mm}
\begin{tabular}{|c|c|c|c|}
\hline
\small  & \small Network Time   & \small 1-second           & \small ACK frame size\\
\small  & \small (ms)           & \small Completion Rate    & \small (Byte)\\
\hline
\small MPNS &\small 104.90 & \small 99.1\% & \small 8.69\\
\hline
\small SPNS &\small 107.68 & \small 99.0\% & \small 6.81\\
\hline
\small Rate &\small +2.65\% & \small -0.10\% & \small -21.63\%\\
\hline
\end{tabular}
\vspace{-2mm}
\end{table}

According to the average ACK frame size of 9 Bytes in MPNS, we set the ack range count limit to 4 in ACK optimization version for SPNS. One drawback was that spurious retransmission could occur more frequently if the number of ACK ranges was limited in a high loss network. Even though in our solution, all received packets is supposed to be acknowledged at least once, the sender may still fail to receive the acknowledgement due to possible ACK packet loss. From Table.2, the average network time of SPNS with 4 ack-range limit increased by more than $2\%$ compared with MPNS. In the meantime, the one-second completion rate dropped by $0.1\%$.

The suppression effect in terms of ACK size was clear. The average size of ACK frame in SPNS was 6.8 Bytes, which was even $21\%$ smaller than that of MPNS.

%% file: discuss.tex
\section{Limitations}\label{sec:discuss}

We have to acknowledge that there are some limitation in the above experiments. We plan to do further testing in the future.

\begin{itemize}[leftmargin=*]

    \item  Small reqeust size in RPC. The data volume transmitted in the Taobao shopping RPC over http scenario was small. When the request size becomes large, we expect the difference to become larger, for example, in video applications.

    \item  Congestion control and path scheduling algorithm. Cubic and minRTT were used in the experiments and we plan to do more testing with BBR.

    \item  0-length CID was not tested. The 0-length CID is currently not supported in this report, so the actual effect of combination of SPNS and 0-length CID is not discussed. 

\end{itemize}

%% file: concl.tex
\section{Conclusion}\label{sec:concl}

We implemented the Multipath extension for QUIC with the multiple packet number space (MPNS) and single packet number space (SPNS) based on the current IETF WG multipath QUIC draft to help the community better understand the implementation complexity and performance tradeoff. The detailed algorithm for SPNS's loss detection, recovery and RTT measurements are discussed in this report. The A/B test results of more than 1.5 million samples in Taobao shopping RPC over HTTP requests showed that SPNS had a slight deterioration in transmission efficiency compared with MPNS. The performance degradation became larger as the request size increased for SPNS. Without ACK size suppression, the acknowledgement cost of SPNS was much higher than that of MPNS. Optimizing the ACK size could work, but may sacrifice the transmission efficiency. Finally, we encourage more testing and experiments from the community to do more validation, and together, help move the multipath QUIC technology forward.